\documentclass[preprint]{aastex}
\usepackage{emulateapj5}
\usepackage{apjfonts}
\usepackage{natbib}

\input psfig.tex

\def\aa{{A\&A}}

\def\aj{{AJ}}
\def\al{$\alpha$}
\def\bet{$\beta$}

\def\annrev{{ARA\&A}}
\def\apj{{ApJ}}
\def\apjs{{ApJS}}
\def\asec{$^{\prime\prime}$}

\def\kms{km s$^{-1}$}
\def\lamb{$\lambda$}
\def\lax{{$\mathrel{\hbox{\rlap{\hbox{\lower4pt\hbox{$\sim$}}}\hbox{$<$}}}$}}
\def\gax{{$\mathrel{\hbox{\rlap{\hbox{\lower4pt\hbox{$\sim$}}}\hbox{$>$}}}$}}
\def\simlt{\lower.5ex\hbox{$\; \buildrel < \over \sim \;$}}
\def\simgt{\lower.5ex\hbox{$\; \buildrel > \over \sim \;$}}
\def\lum{erg s$^{-1}$}

\def\mnras{{MNRAS}}

\def\pasp{{PASP}}

\def\percm2{cm$^{-2}$}
\def\peryr{yr$^{-1}$}

\def\solum{$L_\odot$}
\def\solmass{$M_\odot$}
\def\oii{[\ion{O}{2}]}

\def\hii{\ion{H}{2}}
\def\oiii{[\ion{O}{3}]}

\slugcomment{To appear in {\it The Astrophysical Journal}.}
\shorttitle{[O~{\sc ii}] EMISSION IN QUASARS}
\shortauthors{HO}

\begin{document}

\title{[O~{\sc ii}] Emission in Quasar Host Galaxies: Evidence for a 
Suppressed \\ Star Formation Efficiency}

\author{Luis C. Ho}

\affil{The Observatories of the Carnegie Institution of Washington, 813 Santa 
Barbara St., Pasadena, CA 91101}

\begin{abstract}

The \oii\ \lamb3727 line, a commonly used estimator of star formation rate 
in extragalactic surveys, should be an equally effective tracer of star 
formation in the host galaxies of quasars, whose narrow-line regions are 
expected to produce weak low-ionization emission.  Quasar spectra generally 
show little or no \oii\ emission beyond that expected from the active 
nucleus itself.  The inferred star formation rates in optically selected
quasars are typically below a few \solmass\ \peryr, and some significantly 
less.  Quasars do not appear to occur coevally with starbursts.  Recent 
observations, on the other hand, reveal abundant molecular gas in low-redshift 
quasars.  These two results suggest that the star formation efficiency in 
quasar host galaxies is somehow suppressed during the active phase of the 
nucleus.  The low star formation rates also imply that the nonstellar nucleus 
powers the bulk of the thermal infrared emission in radio-quiet quasars.
\end{abstract}


\section{Introduction}

The discovery of scaling relations between central black hole masses and the
bulge properties of their host galaxies (Magorrian et al. 1998; Gebhardt et al. 
2000; Ferrarese \& Merritt 2000) has stimulated a plethora of ideas 
linking black hole growth with galaxy assembly (see reviews in Ho 2004).  
If the evolution of black holes and galaxies are as closely coupled as 
currently thought, one would expect black hole accretion to show some 
empirical connection with star formation.  Two recent lines of evidence point
in this direction.  From an analysis of a large sample of emission-line 
galaxies selected from the Sloan Digital Sky Survey, Kauffmann et al.  (2003) 
find that narrow-line (Type~2) active galactic nuclei (AGNs) frequently show 
stellar absorption-line features indicative of young to intermediate-age stars, 
with the frequency of young stellar populations growing stronger with 
increasing AGN luminosity.  In a parallel development, studies of quasar 
spectra continue to support the notion that their emission-line regions are 
significantly chemically enriched by episodes of rapid star formation, 
presumably associated with the stellar population in the central regions of 
the host galaxies (Hamann et al.  2004, and references therein).  

It is obviously of considerable interest to measure {\it directly}\ the star 
formation rate (SFR) {\it concurrent}\ with quasar activity.  The existing 
studies on the stellar content of quasar host galaxies have been restricted 
largely to stellar populations belonging to the post-starburst phase (e.g., 
Canalizo \& Stockton 2001; Brotherton et al. 2002; Kauffmann et al. 2003), or 
older (Nolan et al. 2001).  The main challenge in studying the youngest 
stellar population in AGNs is that nearly all the observational tracers 
commonly used to estimate SFRs in galaxies, such as hydrogen recombination 
lines or continuum emission in the ultraviolet, radio, or far-infrared (IR) 
bands, suffer from severe contamination by emission from the AGN itself, if 
the latter is sufficiently strong, as in quasars.  As described in this paper, 
however, there is one important exception.

\oii\ \lamb3727, a prominent nebular emission line in \hii\ regions, is widely 
used to track star formation in galaxy surveys, particularly for redshifts 
$z$ \gax\ 0.4 (e.g., Lilly et al. 1996; Hippelein et al. 2003).  Its use as a 
SFR indicator has been discussed by Gallagher, Bushouse, \& Hunter (1989) and 
Kennicutt (1998), and recently it has been examined extensively by, among 
others, Kewley, Geller, \& Jansen (2004).  Can \oii\ be used to measure SFRs 
in AGNs?  

Now, \oii\ emission is by no means uniquely produced in \hii\ regions.  The 
narrow-line regions of AGNs emit \oii, which can be especially prominent in 
photoionized nebulae characterized by low ionization parameters\footnote{The 
ionization parameter is defined as the ratio of the density of Lyman continuum 
photons to the density of hydrogen.} or in a shock-heated plasma (Ferland \& 
Netzer 1983; Halpern \& Steiner 1983; Ho, Filippenko, \& Sargent 1993a).  On 
the other hand, in narrow-line regions governed by high ionization parameters, 
such as those pertinent to Seyfert galaxies, \oii\ is observed and predicted 
to be relatively weak.  For a plausible range of ionization parameters, 
densities, and ionizing spectra, the intensity of \oii\ \lamb3727 is a small, 
roughly constant fraction of \oiii\ \lamb5007 ($\sim10\%-30\%$), as observed 
(Ferland \& Osterbrock 1986; Ho et al.  1993a; Ho, Shields, \& Filippenko 
1993b).  The 
physical conditions of the narrow-line regions of quasars have been less 
thoroughly studied, but they are thought to be similar to those of Seyfert 
galaxies (Wills et al. 1993).  The approximate constancy of the \oii/\oiii\ 
ratio in high-ionization AGNs (whose ionization parameter can be independently 
gauged by, e.g., the \oiii/H\bet\ ratio), then, suggests a simple strategy for 
estimating SFRs in luminous AGNs such as Seyfert 1 nuclei and quasars:  any 
\oii\ emission in excess of the component intrinsic to the AGN, as constrained 
by the \oiii\ strength, can be reasonably attributed to star formation.

This paper draws attention to the fact that optically selected quasars 
exhibit weak \oii\ emission.  The absence of \oii\ emission in excess of the 
baseline level expected from nonstellar photoionization indicates that 
strong star formation generally does not accompany quasar activity.  
Since nearby quasars harbor significant amounts of molecular gas, their 
low SFRs suggests that nuclear activity somehow curtails the star formation 
efficiency in these systems.  The independent estimate of SFR based on \oii\ 
also helps to clarify the energy source 

\begin{figure*}[t]
\centerline{\psfig{file=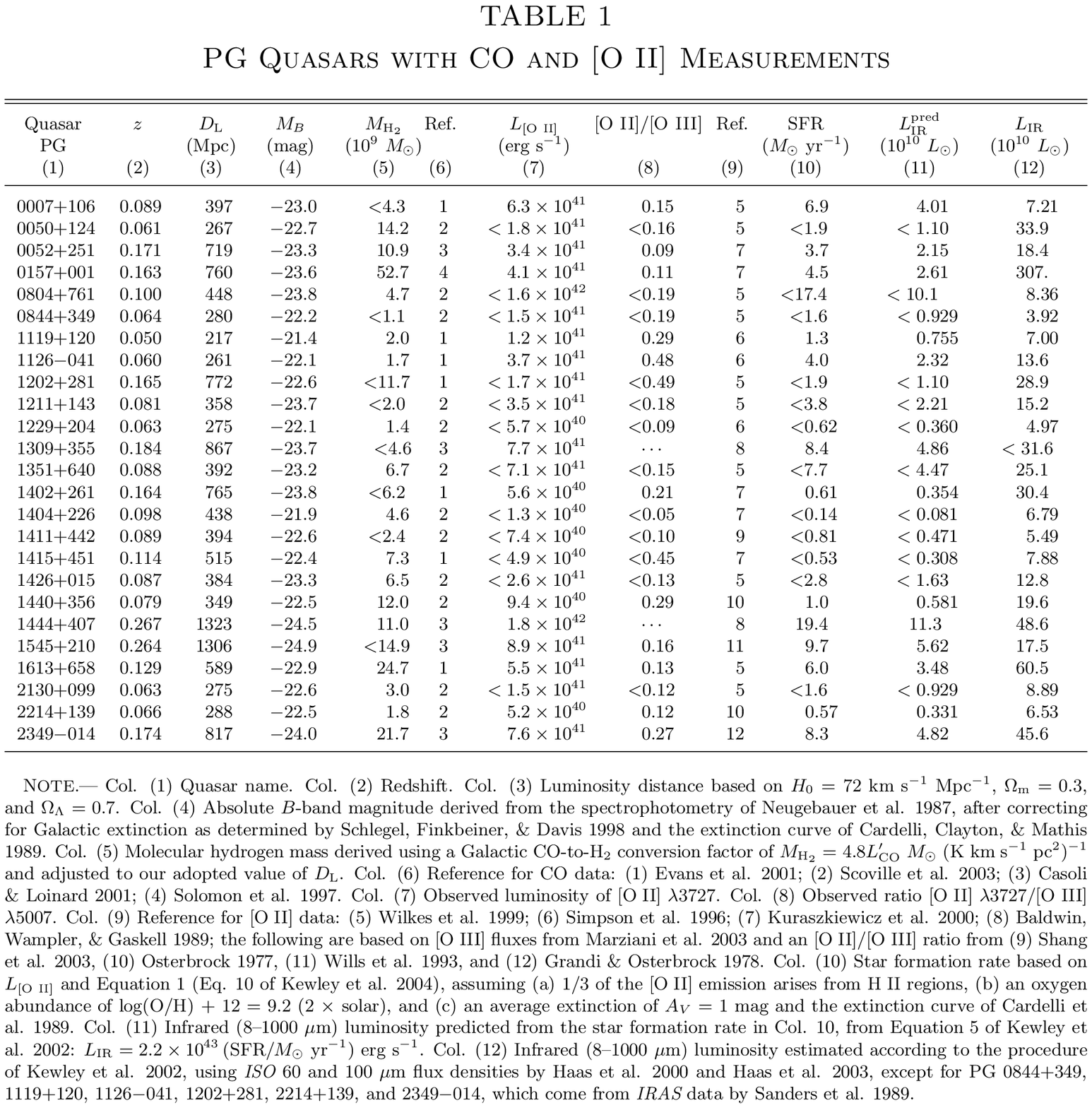,width=18.0cm,angle=0}}
\end{figure*}

\noindent
responsible for the IR emission in quasars.

\vskip +0.3cm
\section{Observational Constraints}

The \oii\ \lamb3727 line has not been extensively studied in quasars.  Unlike 
\oiii\ \lamb5007, which is often prominent in quasars, \oii\ tends to 
be quite weak, in many cases eluding detection.  Radio-loud quasars can have 
\oii\ luminosities as high as $L_{\rm [O~{\sc II}]} \approx 10^{42}-10^{43}$ 
\lum\ (Wills et al. 1993; Hes, Barthel, \& Fosbury 1996).  By contrast, nearby 
radio-quiet quasars from the Palomar-Green (PG) survey (Schmidt \& Green 1983) 
typically have $L_{\rm [O~{\sc II}]} \approx 10^{40}-10^{42}$ \lum, with many 
objects having upper limits near $L_{\rm [O~{\sc II}]} \approx 10^{41}$ \lum\ 
(Simpson et al. 1996; Wilkes et al. 1999; Kuraszkiewicz et al. 2000).  In 
cases where both \oii\ and \oiii\ are observed, \oii/\oiii\ $\approx 0.1-0.3$, 
within the range expected solely from AGN photoionization.  We can arrive at 
a statistically more robust result by examining ensemble averages of large 
quasar samples, as depicted in composite spectra generated from extensive 
 surveys such as the Large Bright Quasar Survey (LBQS; Francis et al. 1991), 
the FIRST Bright Quasar Survey (Brotherton et al. 2001), the Sloan Digital Sky 
Survey (Vanden~Berk et al. 2001; Richards et al. 2003; Yip et al. 2004), and 
the 2dF and 6dF Quasar Redshift Surveys (Croom et al. 2002).  A consistent 
pattern emerges: \oii\ is invariably quite weak (rest-frame EW $\approx$ 2 
\AA), $\sim10\%-20\%$ in strength compared to \oiii.  To estimate the absolute 
strength of \oii, consider the LBQS composite, which comprises mostly of 
quasars with $\langle M_B \rangle \approx -23.5$ mag at 
$\langle z \rangle \approx 1.3$.  For EW(\oii) = 1.9 \AA\ and a continuum 
spectrum $f_\nu \propto \nu^{-0.32}$ (Francis et al. 1991), we find\footnote{
We adopt the following cosmological parameters: $H_0$ = 72 \kms~Mpc$^{-1}$, 
$\Omega_{\rm m} = 0.3$, and $\Omega_{\Lambda} = 0.7$.}
$\langle L_{\rm [O~{\sc II}]} \rangle = 9.8\times 10^{41}$ \lum.  

\vskip 0.3cm

\psfig{file=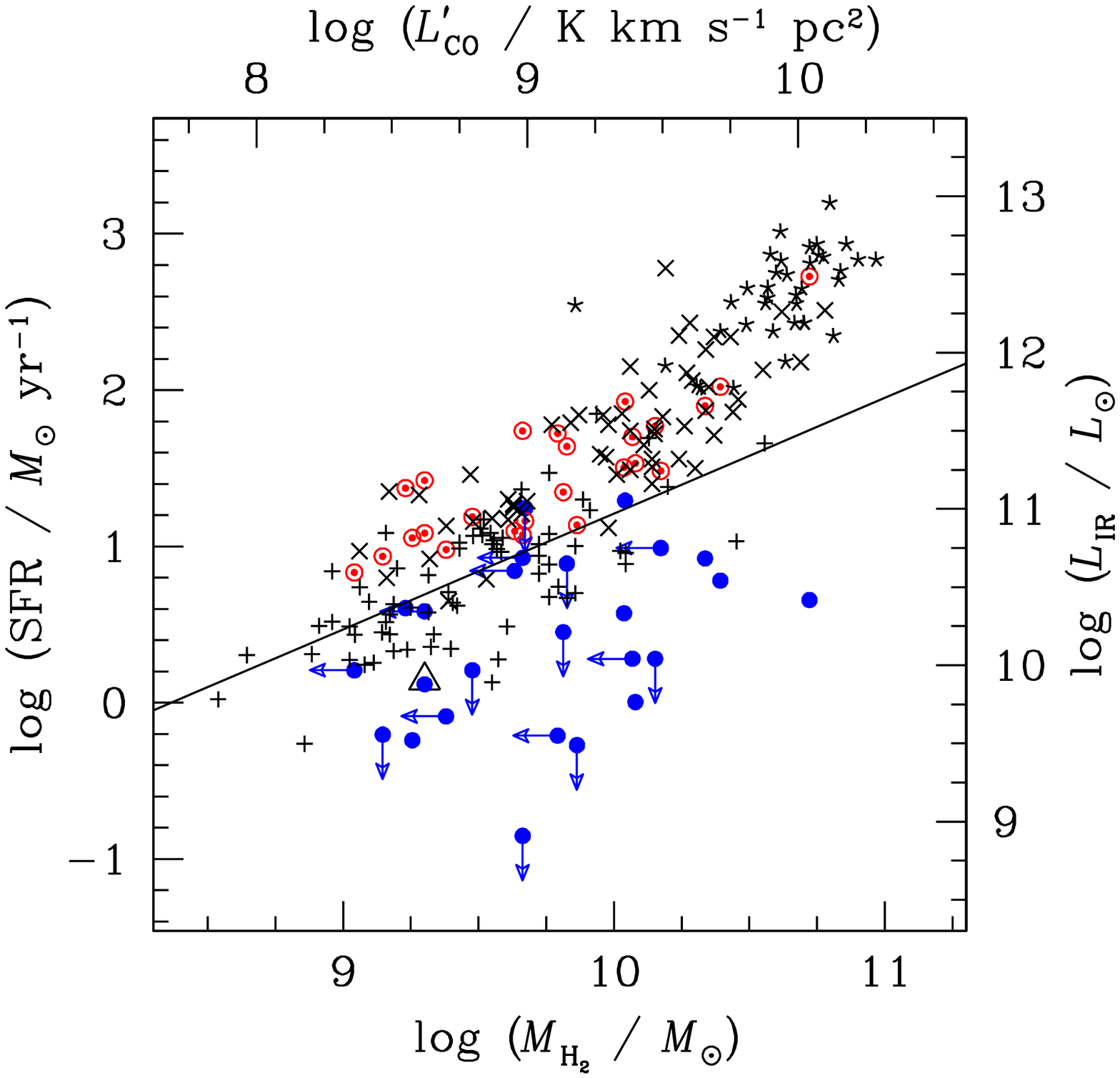,width=8.5cm,angle=0}
\figcaption[fig1.ps]{
The dependence of SFR on molecular gas content in galaxies; the right 
and top axes give the alternative representation in terms of IR luminosity 
and CO luminosity, respectively.  The PG quasars from Table 1 are plotted as 
{\it filled blue circles}, with their SFRs and equivalent IR luminosities 
estimated from \oii\ measurements; upper limits are denoted with arrows.  The 
total IR luminosities actually observed in the PG quasars are indicated with 
{\it semi-filled red circles}.  Three galaxy samples are included for 
comparison, using the prescription of Kewley et al. (2002; see notes to Table 
1) to estimate SFRs from published IR measurements: isolated and weakly 
interacting galaxies ({\it pluses}, with best-fitting line; Solomon \& Sage 
1988), luminous IR galaxies ({\it crosses}; Sanders et al. 1991), and 
ultraluminous IR galaxies ({\it asterisks}; Solomon et al.  1997).  The large 
{\it triangle}\ marks the location of the Milky Way (Scoville \& Good 1989).  
All the literature data have been homogenized using a Galactic CO-to-H$_2$ 
conversion factor of $M_{\rm H_2}$ = 4.8$L^\prime_{\rm CO}$ 
\solmass~(K~\kms~pc$^2$)$^{-1}$ and our adopted cosmological parameters.
\label{fig1}}
\vskip 0.6cm

\section{Implications }
\subsection{Star Formation Rate and Efficiency}

The observations summarized in the preceding section show that \oii\ emission 
in quasars is generically quite weak. When present, its strength relative to 
\oiii\ is entirely consistent with a pure AGN origin, leaving little room for 
any additional contribution from ongoing star formation in the host galaxy. We 
can use the existing \oii\ measurements to place limits on the SFRs\footnote{
Throughout this paper, SFRs refer to a Salpeter stellar initial mass function 
with a lower mass limit of 0.1 \solmass\ and an upper mass limit of 100 
\solmass.}, employing the calibration of Kewley et al. (2004, Eq. 10), which
explicitly takes into account extinction and metalicity corrections:  

\begin{equation}
{\rm SFR([O~{\sc II}])} = \frac{7.9\times10^{-42} (L_{\rm [O~{\sc II}]}/{\rm erg~s^{-1}})}{16.73 - 1.75[\log({\rm O/H})+12]}\ \ \ M_\odot~{\rm yr}^{-1}.  \nonumber
\end{equation}
\vskip 0.3cm

\noindent
Here $L_{\rm [O~{\sc II}]}$ is the extinction-corrected \oii\ luminosity 
and the solar oxygen abundance is assumed to be $\log({\rm O/H})+12$ = 8.9.
We make three assumptions: (1) \oii\ is attenuated on average by $A_V=1$ mag, a 
value commonly deduced in surveys of star-forming galaxies (e.g., Sullivan et 
al. 2000); (2) the line-emitting gas has a characteristic metalicity of 
twice solar, which lies near the upper end of the values inferred for AGN 
narrow-line regions (Storchi-Bergmann et al. 1998); and (3) one-third of the 
\oii\ emission comes from \hii\ regions, which is a conservative upper bound, 
since in actuality the observed \oii/\oiii\ ratios (0.1--0.3) strongly indicate 
that most or all of the \oii\ can be accounted for by the AGN.  Then, an 
observed \oii\ luminosity of $10^{41}$ \lum, a value around which most of the 
current detections and upper limits for PG quasars cluster, translates into a 
SFR of 1 \solmass\ \peryr.  For the LBQS composite spectrum (\S~2), we 
obtain SFR $\approx$ 10 \solmass\ \peryr.  These are interesting limits, 
considering that nearby, normal spiral galaxies (Solomon \& Sage 1988), 
including the Milky Way (Scoville \& Good 1989), have 
SFRs $\approx$ $1-3$ \solmass\ \peryr.

The low SFRs deduced for the quasar host galaxies seem, at first glance, quite 
surprising.  With their nuclei fully shining as powerful AGNs, one would 
na\"{i}vely expect the host galaxies to be quite plentiful in gas, and hence 
forming stars at a rate substantially higher than in quiescent disk galaxies.  
A possible resolution perhaps can be found by appealing to an evolutionary 
scenario, such as that proposed by Sanders et al. (1988), whereby quasars 
emerge as the endpoint of major, gas-rich mergers.  Most of the gas is 
consumed first in a vigorous starburst phase, when the system appears as an 
ultraluminous IR galaxy.  By the time the quasar emerges, the starburst 
has subsided, most of the gas has been exhausted, and any residual gas 
might be expelled by the AGN.

While such a picture has theoretical support from numerical simulations of 
galaxy mergers that include AGN feedback (Springel, Di~Matteo, \& Hernquist 
2005), it appears to be in serious conflict with the observed gas content of 
nearby quasars.  Scoville et al. (2003), in an unbiased survey of optically 
selected (PG) quasars with $z < 0.1$, detected abundant CO(1--0) emission in 
the majority (9/12) of the sources.  With a standard Galactic CO-to-H$_2$ 
conversion factor, the derived molecular gas masses range from $\sim 10^9$ to 
$10^{10}$ \solmass.  Far from being depleted of cold interstellar medium, 
nearby quasars are, in fact, quite {\it gas-rich}.  For comparison, nearby 
disk galaxies contain $M_{\rm H_2} \approx (2-4)\times10^9$ \solmass\ (Solomon 
\& Sage 1988).  Indeed, we are now faced with a new puzzle.  Why do quasar 
host galaxies form so few stars in spite of having significant amounts of 
molecular gas?  To quantify this apparent contradiction, we have examined 25 
low-$z$ PG quasars with available \oii\ and CO data (Table 1)\footnote{Since 
the \oii\ measurements were obtained using apertures ($\sim$2\asec--5\asec) 
that are typically smaller than the optical extent of the host galaxies, the 
quoted \oii\ strengths are probably lower limits to the total line luminosity 
from these objects.  The current comparison with the CO data, however, is valid 
because the CO maps have a resolution of $\sim 4$\asec, and most of the 
emission remains unresolved on this scale (Evans et al. 2001; Scoville et al. 
2003).  Three of the objects (PG~1119+120, 1229+204, and 1404+226) technically 
fall below the traditional luminosity criterion for quasars 
($M_B < -23 - 5\log (H_0/100~{\rm km~s^{-1}~Mpc^{-1}})\,=\,-22.2$ mag for our 
choice of $H_0$), but this historical definition is arbitrary and has no 
particular physical significance.}.  Consistent with the above discussion, the 
majority of the objects have SFRs or limits thereof that are quite modest, 
ranging from less than 1 \solmass\ \peryr\ to a few \solmass\ \peryr.

Gas-rich galaxies ordinarily obey a well-defined correlation between their 
far-IR luminosity and CO luminosity, or, equivalently, between their SFR and 
molecular gas mass.  The ratio of these two quantities yields an estimate 
of the star formation ``efficiency'' (e.g., Young et al.  1986).  This is 
illustrated in Figure 1, where the solid line demarcates the locus occupied by 
isolated or weakly interacting galaxies (Solomon \& Sage 1988). For comparison,
we have superposed a sample of luminous (Sanders, Scoville, \& Soifer 1991) 
and ultraluminous (Solomon et al. 1997) IR galaxies to illustrate the 
familiar result that these objects, most of which are mergers or strongly 
interacting systems, have elevated star formation efficiencies.  In this 
context, it is remarkable that nearly all (23/25) of the PG quasars lie at or
{\it below}\ the locus of normal galaxies.  Recall that the SFRs for the PG 
quasars already should be viewed as conservative upper limits, since 
the \oii/\oiii\ ratios (Table 1) suggest that star formation contributes 
negligibly to the \oii\ emission.  Taken at face value, the host galaxies of 
nearby quasars appear to form stars far less efficiently than would be expected
on the basis of their gas content.  Their star formation efficiencies certainly
fall short of those in strongly interacting galaxies (by a factor $\sim20-50$), 
but, more strikingly, they are suppressed even relative to normal galaxies (by 
a factor \gax 5).

What could be responsible for the anomalously low star formation efficiencies 
in quasar host galaxies?  As reviewed by Maloney (1999), the hard radiation 
field of an AGN can have a profound impact on the thermal and ionization 
structure of a molecular cloud.  Although the details have not been 
elucidated, it is not unreasonable to suppose that this form of AGN feedback
could have direct consequences for a molecular cloud's ability to form stars.  

\subsection{Origin of Far-infrared Emission in Quasars}

Our results have some bearing on the origin of the IR emission in quasars.  
While the IR continuum of quasars, at least of the radio-quiet variety, can be 
explained largely by thermal emission from dust reradiation, the primary 
energy source for the far-IR emission is still controversial (e.g., Haas 
et al. 2003).  The \oii\ measurements discussed above provide an independent 
constraint on this problem.  Table 1 lists the IR luminosity predicted from 
our limits on the SFR, which, as expected, in almost all cases is quite small 
compared to the total observed IR luminosity.  Although the PG quasars 
in Figure 1 indeed do roughly follow the track of IR-luminous galaxies in 
terms of their total IR luminosity, it would be misleading to conclude that 
the IR emission necessarily originates from stellar heating (e.g., Yun et al. 
2004).  The true contribution to the IR emission due to star formation, 
as inferred here through the \oii-based SFRs, push the points well below 
the normal trend.  At least for the subset of PG quasars considered here, it 
seems difficult to escape the conclusion that accretion energy powers the 
bulk of the IR continuum.  

\subsection{Comparison with Previous Work}

The main result of this paper---that quasar host galaxies experience low levels
of ongoing star formation---may appear to be at odds with other work 
in the literature that have inferred high SFRs in quasars.  In a study of 
composite spectra generated from the 2dF+6dF Quasar Redshift Surveys, 
Croom et al. (2002) conclude that the majority of the \oii\ emission in 
low-luminosity ($M_B \approx -20$ mag) objects probably comes from star-forming
regions in the host galaxy and not from the AGN.  These authors speculate that 
the same may hold in high-luminosity sources.  From the observed inverse 
correlation between \oii\ EW and luminosity, and the somewhat steeper 
dependence between Ca~{\sc II}~K absorption EW and luminosity, the assumption 
of a constant host galaxy spectrum would suggest that the AGN contributes an 
increasingly larger fraction of the \oii\ flux in higher-luminosity sources.  
This is consistent with the variation of the relative strengths of \oii\ and 
\oiii\ as a function of luminosity.  From Figure 7 of Croom et al., 
\oii/\oiii\ has a relatively high value of $\sim$0.4 in the low-luminosity 
bin, as expected for significant host galaxy contribution, but decreases to 
$\sim$0.1 at $M_B \approx -24$ mag, a value more typical of AGNs. 
Independent of the exact interpretation, however, we note that the absolute 
strength of \oii\ in the 2dF+6dF composites nonetheless indicates that the 
line luminosity, and thus the SFR, is quite modest.  The high-luminosity 
AGN composite spectra from 2dF+6dF have EW(\oii) $\approx$ 2 \AA\ and span 
$z\approx 1-2$, not unlike the LBQS composite (Francis et al. 1991) discussed 
in \S\S~2--3, and hence our previous limit on the SFR for the latter applies.

Netzer et al. (2004) recently obtained \oiii\ \lamb 5007 measurements of 
luminous, high-redshift quasars in order investigate the properties of their 
narrow-line regions.  From the subset of their sample that shows strong 
\oiii\ emission, Netzer et al. deduced that the narrow-line region in these
objects must be exceptionally dense compared to the conditions in nearby, 
less luminous AGNs.  They postulate that the high-density gas might be related 
to, or supplied by, star-forming regions.  However, this conjecture for the 
origin of the dense gas, while plausible, is not unique, as Netzer et al. 
recognize.  One can envision many possible channels for delivering dense gas 
to the circumnuclear environment of galaxies without implicating nuclear 
starburst activity.  Moreover, even if star formation were ultimately linked 
to the origin of the gas, in the absence of a specific model one cannot 
rule out a scenario wherein the starburst precedes the quasar phase.

As reviewed by Heckman (2004), there has been mounting evidence from
spectroscopic studies of nearby AGNs that nuclear and starburst activity are
closely coupled.  The most extensive treatment of this problem comes from the
analysis of the Sloan Digital Sky Survey database by Kauffmann et al. (2003),
who find that the host galaxies of luminous Type~2 AGNs often show spectral
signatures typical of young to intermediate-age stars.  Although more
difficult to study, broad-line (Type~1) objects qualitatively behave the same
(Kauffmann et al.  2003).  These results are not in conflict with the findings
of this study.  The stellar population uncovered by Kauffmann et al. has a
characteristic age of $10^8-10^9$ yr, which is indicative of a
{\it post}-starburst population, whereas our study specifically aims to
address the younger (\lax $10^7$ yr), ionizing population.

Lastly, there has been considerable success in efforts to detect thermal dust 
emission through sub-millimeter observations of high-redshift ($z$ \gax\ 2) 
quasars (Omont et al. 2004, and references therein). It is customary to invoke 
enormous SFRs, $\sim 10^3$ \solmass\ \peryr\ or more, to account for the large 
far-IR luminosities (\gax $10^{13}$ \solum) observed in these sources, 
especially when CO emission is detected.  While the presence of a sizable 
molecular gas reservoir is clearly a necessary prerequisite for a starburst, 
and undoubtedly indicates the galaxy's future potential to form stars, our 
study shows that this, by itself, is not a sufficient condition for a 
starburst, not when a powerful quasar is simultaneously active.  As discussed 
in \S~3.2 the interpretation of far-IR emission can be ambiguous.  The 
technique outlined in this paper can be applied directly to evaluate the SFRs 
in distant quasars, by searching for \oii\ \lamb3727 emission shifted into the 
near-IR.

\subsection{Caveats}

We conclude with a brief discussion of possible sources of systematic error 
that may affect the results presented in this paper.  First, it is possible 
that we have underestimated the magnitude of dust extinction on the 
\oii\ measurements.  We have assumed that quasar host galaxies on average 
are affected by roughly the same degree of extinction as commonly deduced in 
actively star-forming galaxies.  There is no compelling reason, however, to 
believe that this is a poor approximation.  The amount of dust obscuration in 
galaxies correlates with the level of star formation, but for SFRs \lax\ 100 
\solmass\ \peryr, appropriate for all but the most extreme starbursts,
optical SFR tracers such as \oii\ and H\al\ indicate extinction corrections of 
only a factor of 4--5 (e.g., Dopita et al.  2002; Cardiel et al. 2003; Hopkins 
et al. 2003), which is consistent with our adopted value of $A_V = 1$ mag.  In 
order to explain the apparent offset of the quasars in Figure 1, we would need 
to increase the extinction to $A_V = 2$ mag to match the normal galaxies, and 
to $A_V = 3-4$ mag to be consistent with the IR-bright starbursts.  It is 
worth reiterating that our estimates of SFRs have, if anything, erred on the 
side of caution by assuming that as much as one-third of \oii\ emission arises 
from \hii\ regions, even though the observed \oii/\oiii\ ratios are manifestly 
consistent with an AGN origin.  

It could be objected that the comparison between quasars and star-forming 
galaxies in Figure 1 may have been exaggerated because it invokes two vastly 
different methods of estimating SFRs, namely \oii\ and far-IR luminosity.  
While it would certainly be worthwhile to revisit this analysis once globally 
integrated \oii\ measurements become available for quiescent and ultraluminous 
IR galaxies, here we simply note that the magnitude of the trend observed in 
Figure 1 cannot be explained by possible residual systematic differences 
between the two methods after reasonable precautions are taken to correct 
for extinction in \oii.  The scatter between the SFRs derived from the two 
methods ($\sim 0.25$ dex; Kewley et al. 2002, 2004; Hopkins et al. 
2003) is small compared to the observed scatter in Figure 1.

As emphasized by Kewley et al. (2004), SFRs based on \oii\ can be influenced 
by metalicity.  Although we have no direct information on the metalicity of 
the \oii-emitting gas for our objects, we have made a reasonable guess (twice 
solar) based on detailed studies of nearby AGNs (Storchi-Bergmann et al. 1998) 
and the expectation that the host galaxies of quasars should be relatively 
massive, and hence metal-rich.  Artificially increasing the metalicity to 
3--4 times solar would not qualitatively alter our basic conclusions.  

It is conceivable that quasar host galaxies do experience significantly higher 
SFRs, but somehow most of the \hii\ regions, perhaps as a result of being 
exposed to the strong radiation field of the AGN, are not being properly 
traced through \oii\ emission.  One might also appeal to unusually high 
nebular densities to quench the \oii\ emission by collisional deexcitation.
Both possibilities are purely speculative and difficult to test, but they 
merit further investigation.  

The H$_2$ masses for the quasars were derived using a standard 
CO-to-H$_2$ conversion factor appropriate for Galactic molecular clouds.  
While this is a perennial source of concern, its impact is mitigated 
by the fact that our conclusion concerning star formation efficiency in quasars 
is based on a differential comparison with other galaxies, whose H$_2$ masses 
were derived from the same premise.  Of course, we cannot exclude the 
possibility that quasar host galaxies have a systematically different (lower)
CO-to-H$_2$ conversion factor than other extragalactic systems.  Downes \& 
Solomon (1998) have argued, for example, that the CO-to-H$_2$ conversion 
factor in ultraluminous IR galaxies may be up to a factor of 5 lower than 
the Galactic value.   There is no reason to suspect, however, that the 
extreme conditions in ultraluminous IR galaxies should apply to the quasars 
considered here.  Moreover, the gas masses deduced using the standard 
CO-to-H$_2$ conversion factor, when compared with the dust masses derived 
from IR observations, lead to reasonable gas-to-dust ratios (Haas et al. 2003;
Scoville et al. 2003).  In order to align the quasars with the trend defined 
by the normal galaxies in Figure 1, the H$_2$ masses for the quasars would 
have to be lowered by roughly an order of magnitude.

\vskip 0.5cm
\section{Summary}

This paper discusses the feasibility of using the \oii\ \lamb3727 emission line 
as a tracer of ongoing star formation in AGNs, particularly in the 
high-ionization regime pertinent to Seyfert galaxies and quasars.  From an 
assessment of the existing spectroscopy on quasars, we find that quasars 
exhibit very weak \oii\ emission, with little or no contribution evident from 
star-forming regions in the host galaxy.  Quasar and starburst activity do not 
appear to be coeval.  For a well-defined sample of nearby, optically selected 
quasars with detected CO emission, the low inferred star formation rates 
coupled with the abundant molecular gas suggest that star formation in these 
objects is inefficient, perhaps as a consequence of AGN irradiation.  The low 
star formation rates also imply that the bulk of the IR emission in 
radio-quiet quasars must be powered by accretion energy rather than young 
stars.

\acknowledgements
This work was supported by the Carnegie Institution of Washington and by NASA 
grants from the Space Telescope Science Institute (operated by AURA, Inc., 
under NASA contract NAS5-26555).  The author benefited from discussions with 
K. L. Adelberger, A. J.  Barth, E. J. Barton, J. Darling, E. Keto, and A. 
Oemler.  Z. Shang kindly made available his spectrum of PG~1411+442.  An 
anonymous referee offered constructive comments and suggestions.


%

%


\begin{thebibliography}{}

\bibitem[]{}
Baldwin, J.~A., Wampler, E.~J., \& Gaskell, C.~M. 1989, \apj, 338, 630

\bibitem[]{}
Brotherton, M.~S., Grabelsky, M., Canalizo, G., van Breugel, W., Filippenko, 
A.~V., Croom, S., Boyle, B., \& Shanks, T. 2002, \pasp, 114, 593

\bibitem[]{}
Brotherton, M.~S., Tran, H.~D., Becker, R.~H., Gregg, M.~D., 
Laurent-Muehleisen, S.~A., \& White, R.~L. 2001, \apj, 546, 775

\bibitem[]{}
Canalizo, G., \& Stockton, A. 2001, \apj, 555, 719

\bibitem[]{}
Cardelli, J.~A., Clayton, G.~C., \& Mathis, J.~S. 1989, \apj, 345, 245

\bibitem[]{}
Cardiel, N., Elbaz, D., Schiavon, R.~P., Willmer, C.~N.~A., Koo, D.~C.,  
Phillips, A.~C., \& Gallego, J. 2003, \apj, 584, 76

\bibitem[]{}
Casoli, F., \& Loinard, L. 2001, in Science with the Atacama Large Millimeter
Array, ed. A. Wootten (San Francisco: ASP), 305

\bibitem[]{}
Croom, S.~M., et al. 2002, \mnras, 337, 275

\bibitem[]{}
Dopita, M.~A., Pereira, M., Kewley, L.~J., \& Capaccioli, M. 2002, \apjs, 
143, 47

\bibitem[]{}
Downes, D., \& Solomon, P.~M. 1998, \apj, 507, 615

\bibitem[]{}
Evans, A.~S., Frayer, D.~T., Surace, J.~A., \& Sanders, D.~B. 2001, \aj, 
121, 1893 (erratum: 121, 3285)

\bibitem[]{}
Ferland, G. J., \& Netzer, H. 1983, \apj, 264, 105

\bibitem[]{}
Ferland, G. J., \& Osterbrock, D.~E. 1986, \apj, 300, 658

\bibitem[]{}
Ferrarese, L., \& Merritt, D. 2000, \apj, 539, L9

\bibitem[]{}
Francis, P.~J., Hewett, P.~C., Foltz, C.~B., Chaffee, F.~H., Weymann, R.~J.,
\& Morris, S.~L. 1991, \apj, 373, 465

\bibitem[]{}
Gallagher, J.~S., Bushouse, H., \& Hunter, D.~A. 1989, \aj, 97, 700

\bibitem[]{}
Gebhardt, K., et al.  2000, \apj, 539, L13

\bibitem[]{}
Grandi, S.~A., \& Osterbrock, D.~E. 1978, \apj, 220, 783

\bibitem[]{}
Haas, M., et al. 2003, \aa, 402, 87

\bibitem[]{}
Haas, M., M\"uller, S.~A.~H., Chini, R., Meisenheimer, K., Klaas, U., 
Lemke, D., Kreysa, E., \& Camenzind, M. 2000, \aa, 354, 453

\bibitem[]{}
Halpern, J.~P., \& Steiner, J.~E. 1983, \apj, 269, L37

\bibitem[]{}
Hamann, F., Dietrich, M., Sabra, B. M., \& Warner, C. 2004, in in Carnegie 
Observatories Astrophysics Series, Vol. 4: Origin and Evolution of the 
Elements, ed. A. McWilliam \& M. Rauch (Cambridge: Cambridge Univ. Press), 443

\bibitem[]{}
Heckman, T. M. 2004, in Carnegie Observatories Astrophysics Series, Vol. 1:
Coevolution of Black Holes and Galaxies, ed. L. C. Ho (Cambridge: Cambridge
Univ. Press), 359


\bibitem[]{}
Hes, R., Barthel, P.~D., \& Fosbury, R.~A.~E. 1996, \aa, 313, 423

\bibitem[]{}
Hippelein, H.~H., Maier, C., Meisenheimer, K., Wolf, C., Fried, J.~W.,
von Kuhlmann, B., Kuemmel, M., Phleps, S., \& R\"oser, H.-J. 2003, \aa, 402, 65

\bibitem[]{}
Ho, L.~C. 2004, ed., Carnegie Observatories Astrophysics Series, Vol. 1: 
Coevolution of Black Holes and Galaxies (Cambridge: Cambridge Univ. Press)

\bibitem[]{}
Ho, L.~C., Filippenko, A.~V., \& Sargent, W.~L.~W. 1993, \apj, 417, 63

\bibitem[]{}
Ho, L.~C., Shields, J.~C., \& Filippenko, A.~V. 1993, \apj, 410, 567

\bibitem[]{}
Hopkins, A.~M., et al. 2003, \apj, 599, 971

\bibitem[]{}
Kauffmann, G., et al. 2003, \mnras, 346, 1055

\bibitem[]{}
Kennicutt, R.~C. 1998, \annrev, 36, 189

\bibitem[]{}
Kewley, L.~J., Geller, M.~J., \& Jansen, R.~A. 2004, \aj, 127, 2002

\bibitem[]{}
Kewley, L.~J., Geller, M.~J., Jansen, R.~A., \& Dopita, M.~A. 2002, \apj, 
124, 3135

\bibitem[]{}
Kuraszkiewicz, J., Wilkes, B., Brandt, W.~N., \& Vestergaard, M. 2000, \apj,
542, 631

\bibitem[]{}
Lilly, S.~J., Le F\`evre, O., Hammer, F., \& Crampton, D. 1996, \apj, 460, L1

\bibitem[]{}
Magorrian, J., et al.  1998, \aj, 115, 2285

\bibitem[]{}
Maloney, P. R. 1999, Ap\&SS, 266, 207

\bibitem[]{}
Marziani, P., Sulentic, J.~W., Zamorani, R., Calvani, M., Dultzin-Hacyan, 
D., Bachev, R., \& Zwitter, T. 2003, \apjs, 145, 199

\bibitem[]{}
Netzer, H., Shemmer, O., Maiolino, R., Oliva, E., Croom, S., Corbett, E.,
\& di~Fabrizio, L. 2004, \apj, 614, 558

\bibitem[]{}
Neugebauer, G., Green, R.~F., Matthews, K., Schmidt, M., Soifer, B.~T., \& 
Bennett, J. 1987, \apjs, 63, 615

\bibitem[]{}
Nolan, L.~A., Dunlop, J.~S., Kukula, M.~J., Hughes, D.~H., Boroson, T., 
\& Jimenez, R. 2001, \mnras, 323, 308

\bibitem[]{}
Omont, A., Beelen, A., Bertoldi, F., Carilli, C. L., \& Cox, P. 2004, in 
Multiwavelength AGN Surveys, ed. R. M\'ujica \& R. Maiolino (Singapore: 
World Scientific), 109

\bibitem[]{}
Osterbrock, D.~E. 1977, \apj, 215, 733

\bibitem[]{}
Richards, G.~T., et al. 2003, \aj, 126, 1131

\bibitem[]{}
Sanders, D.~B., Phinney, E.~S., Neugebauer, G., Soifer, B.~T., \& Matthews, K. 
1989, \apj, 347, 29

\bibitem[]{}
Sanders, D.~B., Scoville, N. Z., \& Soifer, B.~T. 1991, \apj, 370, 158

\bibitem[]{}
Sanders, D.~B., Soifer, B.~T., Elias, J.~H., Madore, B.~F., Matthews, K.,
Neugebauer, G., \& Scoville, N.~Z. 1988, \apj, 325, 74

\bibitem[]{}
Schlegel, D.~J., Finkbeiner, D.~P., \& Davis, M. 1998, \apj, 500, 525

\bibitem[]{}
Schmidt, M., \& Green, R.~F. 1983, \apj, 269, 352

\bibitem[]{}
Scoville, N.~Z., Frayer, D.~T., Schinnerer, E., \& Christopher, M. 2003,
\apj, 585, L105

\bibitem[]{}
Scoville, N.~Z., \& Good, J. C. 1989, \apj, 339, 149

\bibitem[]{}
Shang, Z., Wills, B.~J., Robinson, E.~L., Wills, A., Laor, A., Xie, B., \& 
Yuan, J. 2003, \apj, 586, 52

\bibitem[]{}
Simpson, C., Ward, M.~J., Clements, D.~L., \& Rawlings, S. 1996, \mnras, 281, 
509

\bibitem[]{}
Solomon, P.~M., Downes, D., Radford, S.~J.~E., \& Barrett, J.~W. 1997, \apj, 
478, 144

\bibitem[]{}
Solomon, P.~M., \& Sage, L.~J. 1988, \apj, 334, 613

\bibitem[]{}
Springel, V., Di Matteo, T., \& Hernquist, L. 2005, \mnras, in press

\bibitem[]{}
Storchi-Bergmann, T., Schmitt, H.~R., Calzetti, D., \& Kinney, A.~L. 1998, 
\aj, 115, 909

\bibitem[]{}
Sullivan, M., Treyer, M.~A., Ellis, R.~S., Bridges, T.~J., Milliard, B., \& 
Donas, J. 2000, \mnras, 312, 442

\bibitem[]{}
Vanden Berk, D.~E., et al. 2001, \aj, 122, 549

\bibitem[]{}
Wilkes, B.~J., Kuraszkiewicz, J., Green, P.~J., Mathur, S., \& McDowell,
J.~C. 1999, \apj, 513, 76

\bibitem[]{}
Wills, B.~J., Netzer, H., Brotherton, M.~S., Han, M., Wills, D., Baldwin, 
J.~A., Ferland, G.~J., \& Browne, I.~W.~A. 1993, \apj, 410, 534

\bibitem[]{}
Yip, C.~W., et al. 2004, \aj, 128, 2603

\bibitem[]{}
Young, J.~S., Kenney, J.~D., Tacconi, L., Claussen, M.~J., Huang, Y.-L., 
Tacconi-Garman, L., Xie, S., \& Schloerb, F.~P. 1986, \apj, 311, L17

\bibitem[]{}
Yun, M.~S., Reddy, N.~A., Scoville, N.~Z., Frayer, D.~T., Robson, E.~I., \&
Tilanus, R.~P.~J. 2004, \apj, 601, 723

\end{thebibliography}
\end{document}